# Glass-like Electronics in Vanadium Dioxide

M. Samizadeh Nikoo[1,*], R. Soleimanzadeh[1], A. Krammer[2], Y. Park[3], J. Son[3], A. Schueler[2], P. Moll[4], and E. Matioli[1,*]

[1]Electrical Engineering Institute, École Polytechnique Fédérale de Lausanne (EPFL), Lausanne, Switzerland.

[2]Institute of Physics, École Polytechnique Fédérale de Lausanne (EPFL), Lausanne, Switzerland.

[3]Department of Materials Science and Engineering (MSE), Pohang University of Science and Technology (POSTECH), Pohang, Republic of Korea.

[4]Institute of Materials Science and Engineering, École Polytechnique Fédérale de Lausanne (EPFL), Lausanne, Switzerland.

*mohammad.samizadeh@epfl.ch, elison.matioli@epfl.ch

**Functional materials can offer new paradigms for miniaturized and energy-efficient electronics, providing a complementary or even alternative platform to metal-oxide-semiconductors. Here we report on electronically accessible long-lived structural states in Vanadium Dioxide that can offer a scheme for data storage and processing. We show that such states can be electrically manipulated and tracked beyond 10,000 seconds after excitation, exhibiting similar features of glasses, which are beyond the classic metastability in Mott systems. Glass-like electronics can potentially overcome some of the fundamental limitations in conventional metal-oxide-semiconductor electronics, and open avenues for neuromorphic computation and multi-level memories.**

The electrostatic control in metal-oxide-semiconductor junctions has provided a platform to achieve a variety of functionalities in an integrated circuit form factor, which constitute the building block of today's electronics[1-3]. This powerful technology, however, faces fundamental constraints for further miniaturization and reduction of energy consumption, which has led to a significant body of research on new materials, devices, integration technologies, and architectures for storage and computation[4]. For example, memristors can enable extremely high density memory platforms, much higher than that of transistor-based memories with identical lithography nodes[5,6], however, as non-volatile resistive switches their operation is limited to specific applications.

Strongly-correlated materials, in which several physical interactions involving spin, charge, lattice, and orbital are simultaneously active, display notable electrical properties[7]. Among them, the first-order insulator-metal transition (IMT) in vanadium dioxide ($VO_2$) happening close to room temperature has attracted considerable interest[8-13]. From a physical point of view, understanding the underlying mechanism of phase switching in $VO_2$ is still a challenge in condensed matter physics, as several models ranging from Peierls- to Mott-Hubbard-type were not successful in explaining the broad range of phenomena occurring in the material[14]. Different types of excitations such as temperature, electric field, and doping can induce the IMT, which makes the phase switching more challenging[15]. From a technological point of view, the bulk conductivity and abrupt phase transition in $VO_2$ can potentially overcome some of the fundamental limitations in conventional metal-oxide-semiconductor electronics, such as the limited conductance imposed by Thomas–Fermi screening[16] and the thermionic subthreshold-slope limit imposed by Boltzmann–tyranny[17].

Here we demonstrate yet another exotic property of Vanadium Dioxide — glass-like dynamics that can be electrically probed and manipulated — which can offer a platform for information processing and storage. We show that a two-terminal device exhibits a continuous spectrum of states that can be revealed by the incubation time of the IMT: the time at which the nucleation of phase transition percolates to form the first conductive filament between two terminals of the switch. The state can be excited by a sequence of binary switching events and can be tracked over $10^4$ s after the excitation. The system exhibits features of a glassy state, presenting a memory effect much longer than that described by classic metastability in Mott insulators.

Fig. 1a shows an ultrahigh sample-rate time-domain experimental setup that can precisely collect temporal response of a $VO_2$ switch (inset). The device was integrated with radiofrequency pads (ground-signal-ground configuration) which together with high-frequency probes enable precise measurements with time resolutions down to ~5-ps[18]. A square pulse generator applies repetitive 10-μs-long pulses with a fixed amplitude (set voltage $V_{set}$ = 2.1 V) to a two-port 3-μm long $VO_2$ switch. The waveform of the current passing through the device is measured for each pulse at the 50-Ω port of a high-frequency oscilloscope, and the transient conductance of the device is extracted. Following an applied pulse, the $VO_2$ film exhibits initially an insulating behavior, and only after an incubation time $t_{inc}$, it undergoes IMT (Fig. 1b). The measurements indicate that the incubation time strongly depends on the history of the previous phase transitions. The very first switching curve presented in Fig. 1b shows an incubation time of ~1.4 $\mu$s. Triggering an IMT and measuring the incubation time after a 10-ms-long relaxation time ($T$) results in a 10 times shorter incubation time. Longer relaxation times after the first phase transition cause longer incubation times, however, the value of $t_{inc}$ is still lower than that of the very first switching, even after $T$ = 10,000 s.

Incubation time versus relaxation time, shown in Fig. 1c, indicates a logarithmic relation $t_{inc} = t_{inc0} \log(T/T_0)$, with $t_{inc0}$ = 78 ns and $T_0$ = 160 $\mu$s. Although $t_{inc}$ has a strong dependence on the previous switching events, the device conductance $G_{ins}$ in the insulating-state (averaged over 20 ns < $t$ < 120 ns), shows very small variations that become undetectable after ~1 s (Fig. 1d). The observed effect is qualitatively identical in micrometer- and nanometer-long devices, enabling potential technological uses of this effect in a miniaturized form factor. Fig. 1e shows incubation time versus the relative increase in conductivity for a 100-nm-long channel device. The values of incubation times are well distinguishable while the conductivity shows a very small variation, which completely relaxes after 1 second. The thermal relaxation of the device is also quite fast (~100 ns, see Fig. S1), and therefore, the heat accumulation does not play any role in these observations (inset of Fig. 1e). The memory effect observed in $t_{inc}$ is reversible and is not due to any degradation in the film (Fig. 1f). The here-presented results correspond to a 100-nm-thick film $VO_2$ synthesized by sputtering on a high resistivity silicon substrate (Fig. S2), however, the results were reproduced in PLD-grown films on other substrates, showing the generality of the effect (Fig. S3).

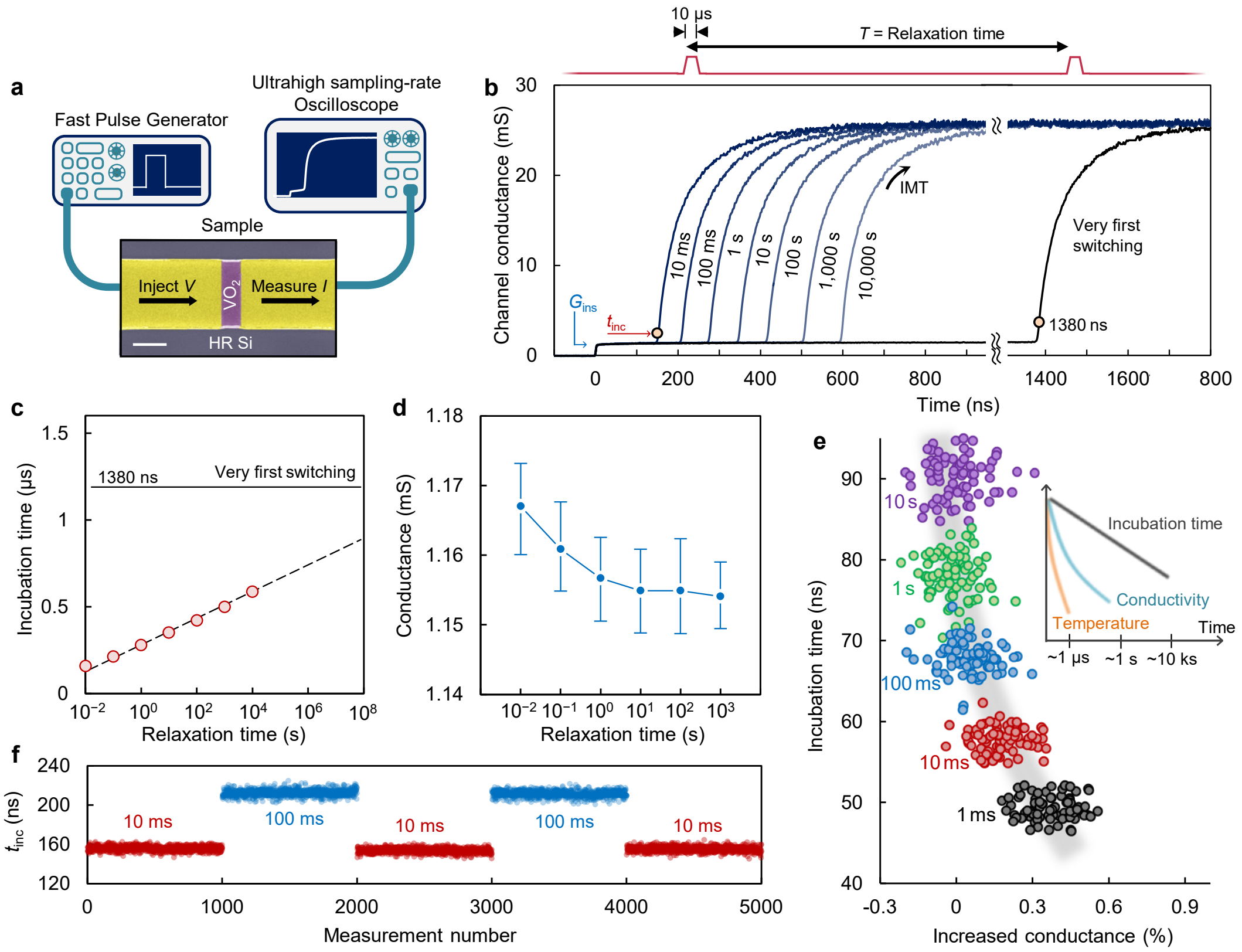

**Fig. 1. Tracing state dynamics of $VO_2$ switches with incubation time.** (**a**) Schematic of the experimental ultrafast time domain setup. The SEM image shows the $VO_2$ switch (scale bar corresponds to 5 μm). Here the devices investigated had lengths varying from 50 nm to 3 μm. (**b**) Transient conductance of the $VO_2$ channel corresponding to different relaxation times $T$, as well as the very first switching cycle. Incubation time ($t_{inc}$) corresponds to the interval between the applied pulse and the onset of the change in conductance due to IMT. The insulating-state conductance $G_{ins}$ corresponds to the average of conductance in the time interval between 20 ns and 120 ns. (**c**) Incubation time versus relaxation time. $t_{inc}$ shows a logarithmic function of $T$. The error bars are smaller than the point dimension. (**d**) Conductance of the insulating state versus relaxation time. After ~1 s, variations in the conductance can no longer be detectable. (**e**) Incubation time versus increased conductance ($(G_{\mathrm{ins}} - \bar{G})/G_{\mathrm{ins}}$, where $\bar{G}$ is the average over $G_{ins}$ for measurements corresponding to 10 s relaxation) for different relaxation times between 1 ms and 10 s for a 100-nm-long device, showing the reproducibility of the results. The inset illustrate the fast relaxation of temperature and resistance, and the slow dynamics of incubation time. (**f**) Monitored incubation time for 5,000 measurements with two different relaxation times 10 ms and 100 ms, showing that the effect is reversible and consistent.

In a following experiment, the devices were excited with a packet of $N$ identical pulses ($V_{set}$ = 2.1 V) and $t_{inc}$ was monitored after a relaxation time $T$ = 1 s (Fig. 2a). The total duration of the excitation including $N$ pulses ($\Delta T$ = 1 ms) was much shorter than the relaxation time $T$. Different pulse packets result in different values of $t_{inc}$, indicating that such multi-pulse excitation can be used to manipulate the state of the device. Fig. 2b shows the summary of the observed results in which the measured values of $t_{inc}$ for different values of $N$ are presented for 60 consecutive acquisitions. These results indicate a fully-electrical scheme to manipulate and sense a Mott system (Fig. 2c). It is worth noting that the excited state does not depend on $\Delta T$, and so the state can be induced in very short time scales. Incubation time in the evaluated devices also did not show any dependence on the width of excitation pulses and the state manipulation was achieved only through the number of triggering events.

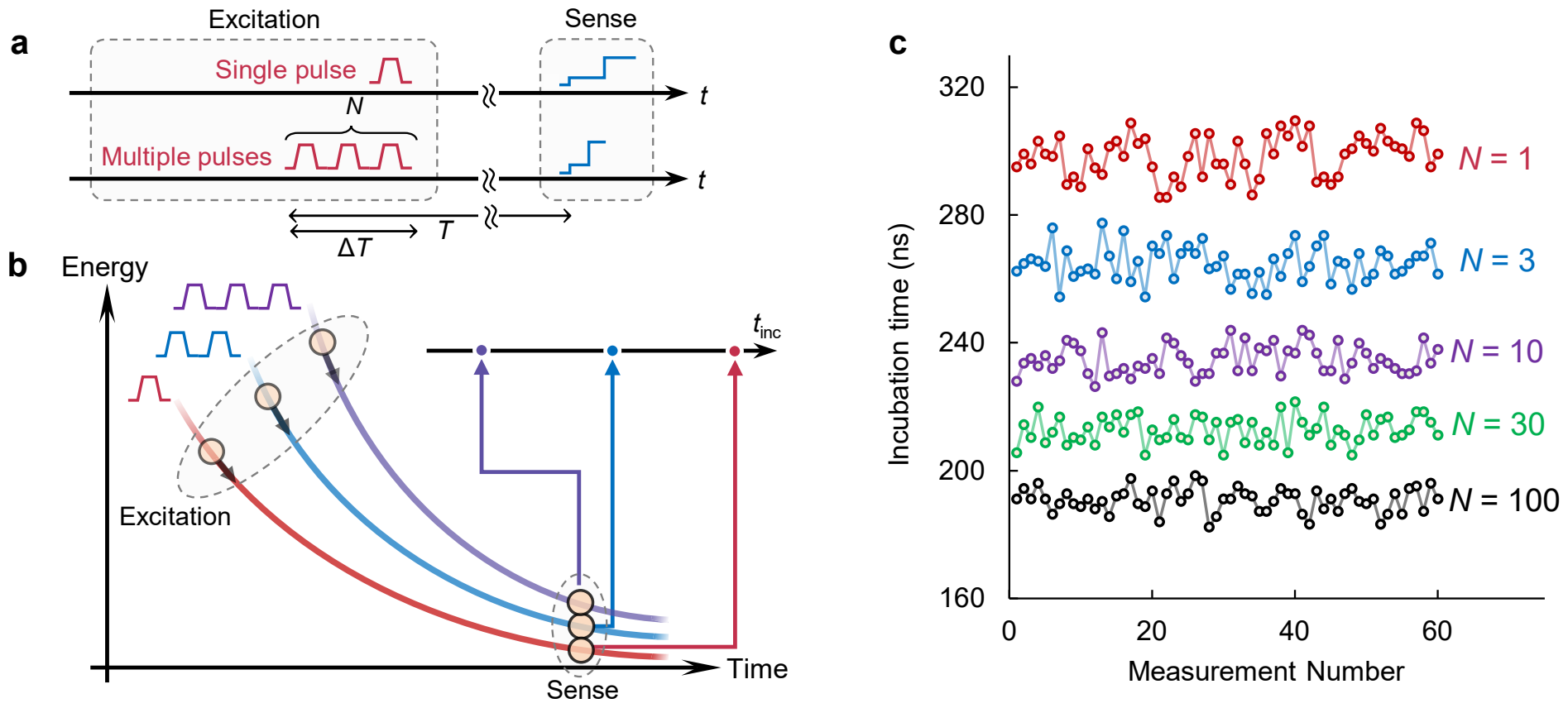

**Fig. 2. Manipulation of post-firing state of $VO_2$ switches.** (**a**) Schematic showing excitation of the $VO_2$ switch with single or multiple pulses and monitoring of the $t_{inc}$ at the sensing pulse after a long relaxation time. (**b**) After multiple pulses, the film undergoes different relaxation trajectories, and monitoring $t_{inc}$ can well identify the different original excitations. (**c**) Measured $t_{inc}$ for different excitation pulse numbers ($N$) for $\Delta T$ = 1 ms and $T$ = 1 s.

We explored the possible mechanisms of the observed effect. The first question was whether the electric current in the film induced the memory. This is a critical aspect because quasi-non-volatile memories induced by long high-current biases have been observed[19]. However this is not the case in our device, as the memory effect is independent from the excitation type and can also be induced purely by heat. Fig. 3a shows synchronized measurements in which IMT is achieved by a pulsed heater that is electrically isolated from the $VO_2$ switch. Every two seconds, a 20-$\mu$s-long pulse triggers IMT in the electrically isolated heaters, which ensures a temperature-driven IMT in the $VO_2$ switch at the middle of two heaters (see Fig. S4). We electrically triggered the middle $VO_2$ switch with a period of one second and monitored the incubation time. The electrical excitation has a 20-ms time lag with respect to the heater signal (Fig. 3b). Therefore, in one electrical measurement, the $VO_2$ switch had a 20-ms relaxation time after a thermal-IMT, and in the next electrical measurement, the switch had a much longer relaxation (1 second) which is considered to be the reference measurement (without memory). The results indicate that the thermally-driven IMT also induces a change in incubation time (Fig. 3c). This suggests the generality of the effect, and at the same time shows that the electric current in the film, which could induce excitation or movement of ions, does not play any role in the memory effect since there was no current flowing in the device during the thermally-driven IMT. In addition, our measurements show that the effect is independent from metal-$VO_2$ junction (Fig. S5), and so the memory is solely due to the intrinsic relaxation of $VO_2$ film after the metal-insulator transition.

A possible mechanism for memory effect in $VO_2$ is the transient metastable states in the metal-to-insulator path[20,21]. However, these nonequilibrium states that exist in form of monoclinic metallic domains are unlikely responsible for our observation, for the following reasons: First, the relaxation of metastable phases is known to be in the range of milliseconds which is orders of magnitude shorter than the memory duration observed[22]. Second, our measurements indicate different incubation times even when the resistance relax to the steady state (Fig. 1d and 1e). Finally, devices with very different sizes, from 50 nm to a few micrometers, exhibit similar behaviors, showing that the memory effect cannot be attributed to mesoscopic scale phenomena.

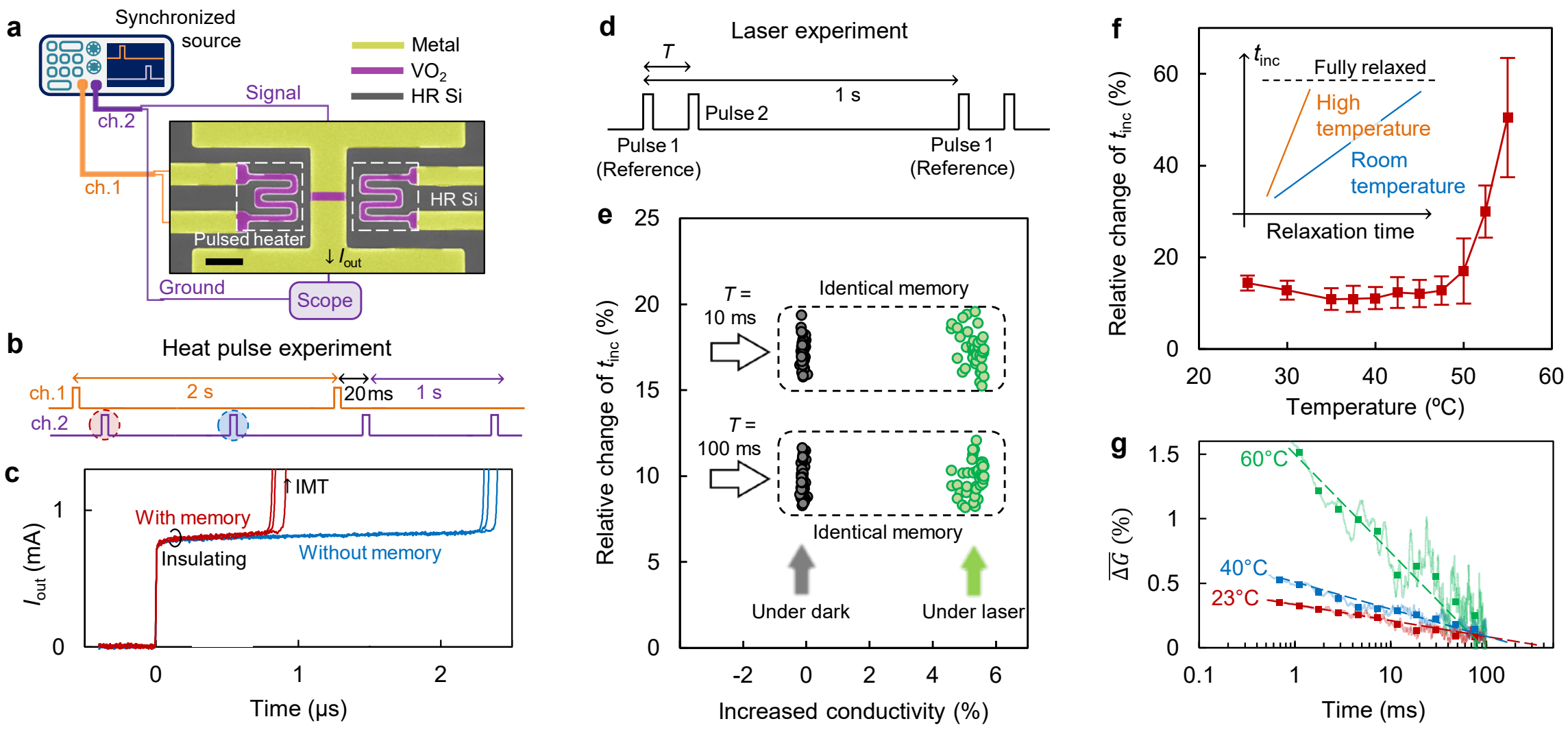

**Fig. 3. Exploration on possible mechanisms behind the long memory.** (**a**) Schematic of an experimental setup for a synchronized measurement in which the memory is induced by a heat pulse, without flowing current through the $VO_2$ device. (**b**) Illustration of waveforms in the synchronized measurement. Channel 1 triggers the IMT in electrically-insulated pulsed heaters which ensure thermal-driven phase transition in the middle switch. 20-ms after thermal IMT, the middle switch is triggered by a voltage pulse and the incubation time (dashed red circle) is compared with the case of no thermal-driven IMT (dashed blue circle). After 20 ms, the switch was back to room temperature, because of the very small excitation duty cycle (0.1%), which is also reflected by the similar insulating state conductivities with and without memory effect. (**c**) Measurements for three consecutive cycles indicate that the memory can be induced by thermal IMT, showing the generality of the effect. (**d**) Illustration of excitation signal of a $VO_2$ switch for measurements under laser light. Every second, two identical pulses with the time separation $T$ are applied to the device and incubation times are monitored ($t_{\rm inc}^{\rm ref}$ and $t_{\rm inc}$, respectively). (**e**) Relative change in the incubation time ($(t_{\rm inc} - t_{\rm inc}^{\rm ref})/t_{\rm inc}^{\rm ref}$) versus increased insulating-state conductivity in a 200-nm-long channel $VO_2$ switch. Measurements under a continuous-wave 532-nm laser light with ~100 W cm$^{-2}$ power density show a considerably higher conductivity, however, the memory effect is unchanged for both different relaxations times $T$ = 10 ms and 100 ms. (**f**) Relative change in the incubation time after $T$ = 10 ms with respect to reference pulses with relaxation time of $T$ = 1 s at different temperatures. There more pronounced change in $t_{\rm inc}$ when the temperature becomes close to the IMT temperature indicates a faster relaxation (inset). (**g**) Increased conductance of the device after IMT. Average over 200 measurements (solid lines), average over exponentially separated time steps (solid points), and the linear interpolations (dashed lines). Monitoring the conductivity also indicates a faster relaxation at elevated temperatures.

We examined more generally whether electronic states could be responsible for our observation. We evaluated the memory effect under a continuous-wave (CW) laser pumping with relatively high energy photons (2.33 eV), larger than the material band gap[23]. The measurements under laser excitation show a considerably higher conductivity with respect to those under dark, however, the memory effect is not affected (Fig. 3d and 3e). This observation does not support scenarios that explain the memory effect based on excitation of electronic states, such as 3d orbital occupancy or trapped carrier in defects. In fact, logarithmic or stretched exponential relaxations are quite slow processes for electronic states, however, such long relaxation times are one of the main features of glassy states with configurational transitions[24]. The IMT in $VO_2$ is a structural switching, and therefore, slow glass-like configurational changes driven by bond lengths or vacancies can be expected.

To investigate this possible scenario, we evaluated the memory effect at elevated temperatures to examine one important feature of a glass relaxation: the relaxation process becomes faster close to the glass transition temperature[25]. In a first experiment, we monitored the change of incubation time corresponding to relaxation times of 10 ms and 1 second. This difference is proportional to the slope of the dashed line shown in Fig. 1c. At each temperature set point, we set the voltage such that the incubation time in the pulse with 1 second relaxation remained fixed (at 500 ns). The measurement results shown in Fig. 3f indicates that the sensitivity of incubation time with respect to the relaxation time is substantially increased close to the IMT temperature. This means a steeper slope for $t_{inc}$-$T$ curve which shows that the system requires a shorter time to get fully relaxed (inset of Fig. 3f). In a second experiment, we monitored the temporal change in the device conductance after an IMT event. The measurement results shown in Fig. 3g also indicates that the relaxation of the device conductance becomes faster at elevated temperatures. As a result, both Figures 3f and 3g show a more rapid relaxation close to the IMT, which is a feature of a glass relaxation. Glass transitions are also reversible and can undergo new phase paths if they are heated before a complete relaxation. These are analogous features of our observations shown in Fig. 1f and Fig. 2b, respectively. It should be noted that identical memory effect was obtained under vacuum of $10^{-4}$ mbar, which shows that such off-stoichiometric mechanism must be purely configurational, without any exchange with the ambient. Even though more investigations are needed to identify the microscopic details of this process, the observed effect certainly shares several important features of glasses, thus can be considered as glass-like dynamics.

The concept of a glass-like state that is electronically accessible can offer opportunities in electronics. The very slow relaxation together with the arbitrary manipulation capability is of interest for data storage, especially since the two-terminal configuration of the devices can be easily scaled down, and operate with very fast dynamics and low energy (Fig. S6). It is also compatible with a cross-bar configuration enabling extremely high data-storage densities[5]. Another important aspect of the observed effect is that, at constant amplitude, $t_D$ is proportional to the energy required to activate IMT. Therefore, the history of a device determines the switching energy barrier ($E_{bar}$) in the future: the higher the number and frequency of switching events, the lower the energy barrier (Fig. 4a and 4b). One can model $E_{bar}$ at $t = t_0$ in the transient regime after $n$ pulses occurring at $t_k$ ($1 \leq k \leq n$) by

$$E_{bar} = E_0 - E_1 \ln\left(\sum_{k=1}^{n} \frac{\tau}{t_0 - t_k}\right), \tag{1}$$

in which the second term in the right hand of the equation represents the reduced energy barrier due to the film history. This functionality can enable highly dynamic classifiers with a computation-free training, which cannot be achieved in classic classifiers based on nonlinear resistive elements[26]. To show this, we can consider a neural network with one hidden layer, in which a $VO_2$ switch has been placed between each two nodes (Fig. 4c). At time $t$, the network can be fully described by matrices $E_{bar}^{A}(i, j, t)$ and $E_{bar}^{A}(j, k, t)$ representing the reduced energy barrier of $A_i$-$X_j$ and $X_j$-$B_k$ switches, respectively. The product of these two matrices represents the correlation between inputs and outputs, enabling an energy-based classification: for each set of inputs, the output with the minimum energy required for IMT triggering at the interconnections will be activated.

This concept provides two important features. First, training of the network can be done purely based on hardware. There is no need for calculation of weights and also no need to physically induce them, for example to manipulate the resistivity of elements[26]. We show this feature in classification of three characters “I”, “J”, and “L” provided in $3 \times 3$ pixels (Fig. 4d). Application of electric currents to the input nodes ($A_i$) corresponding to each image label, and grounded the equivalent output, can simply train the network. This reduces the energy barrier for some of the pathways connecting each sets of inputs to the corresponding output. Fig. 4e shows the training result of the network, which results in successful classification of characters (Fig. 4f). The second feature provided by a glassy neural network is its high dynamics. The system is operational after an initial training, however, it fine tunes itself as it is exposed to the classification samples, because each sample would induce switching events at the correct nodes, reducing their energy barriers. This can self-optimize the network during the classification process, only relying on a small training set.

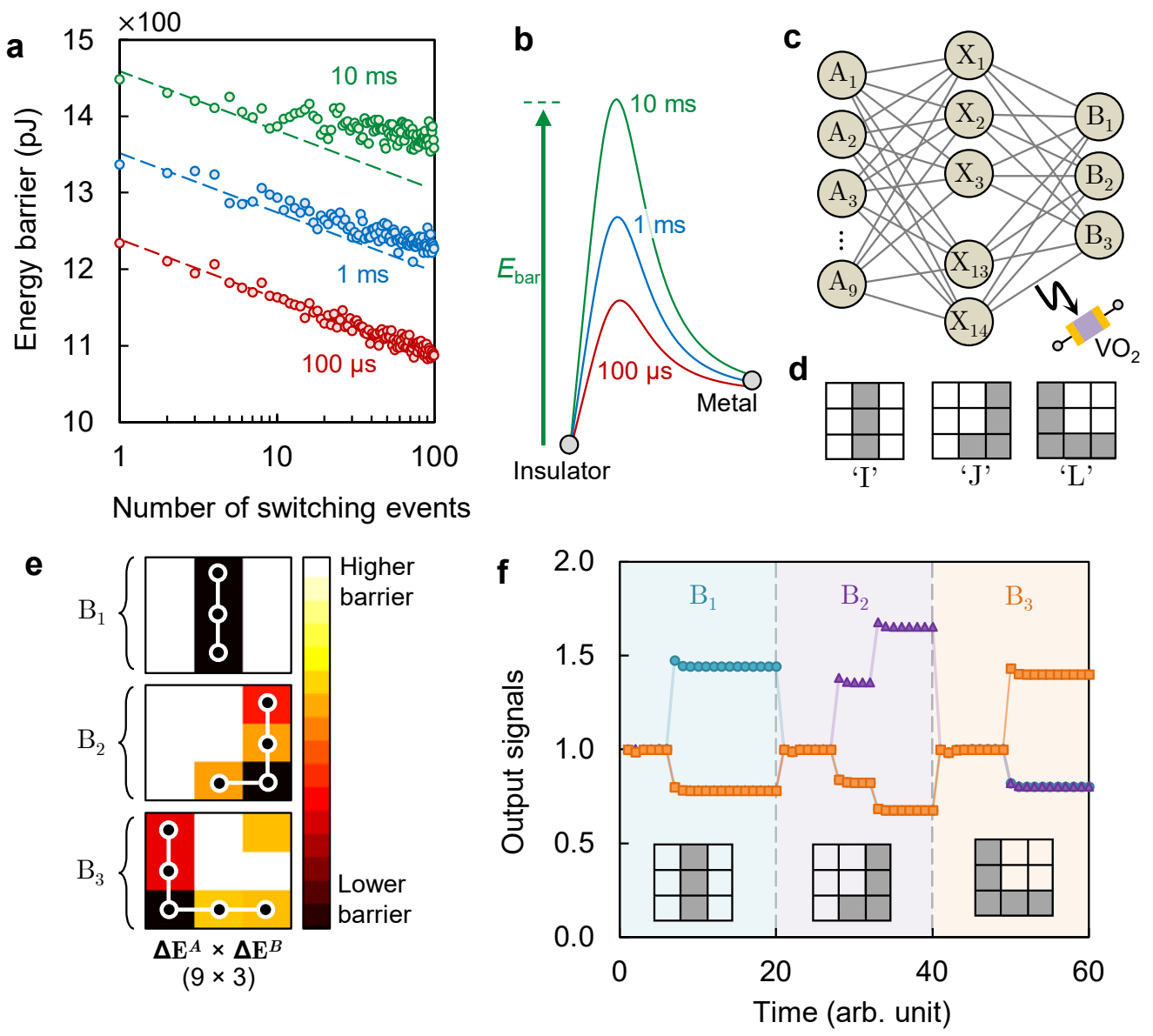

**Fig. 4. Glass-like electronics for highly-dynamic brain-inspired computation.** (**a**) Dynamic change in energy barrier ($E_{bar}$) of a 2-μm-long $VO_2$ switch with the number of switching events for different pulse separations. (**b**) Schematic showing that shorter pulse separations result in a lower $E_{bar}$. (**c**) Considered neural network. A $VO_2$ switch is placed in the interconnection between each two nodes. $\Delta E^A$ and $\Delta E^B$ matrices represent the reduced energy barriers corresponding to $A_i$-$X_j$ and $X_j$-$B_k$ interconnections. (**d**) Labels used to train a 3 × 3 pixel network for image classification. (**e**) Post-training $\Delta E^A \times \Delta E^B$ matrix represents the effective reduced energy barrier corresponding to each output. (**f**) Output signals at the three nodes $B_1$ (circles), $B_2$ (triangles), and $B_3$ (squares). Exposing the network to the inputs and monitoring the output currents successfully classifies the samples.

As a conclusion, our work demonstrates glass-like phase evolutions in $VO_2$ that can be induced in sub-nanosecond time scales and monitored during several orders of magnitudes in time, from sub millisecond to above one hour. A two-terminal switch undergoes a complex but fully reversible phase dynamics, induced by a series of excitations. In a technological point of view, our results show that the sensitivity of the phase dynamics to the number of excitation pulses and their time interval can enable all-electrical integrated schemes for data storage and processing. From a physical point of view, our work revealed extremely long memories in $VO_2$ that can be only revealed by monitoring the incubation time. This can set the stage to study the dynamics of out-of-equilibrium phases in other material systems[27,28]. More generally, glass-like electronics can potentially overcome some of the fundamental limitations in conventional metal-oxide-semiconductor electronics such as the continuous demand of decreased voltage supply levels[29,30], and open avenues for novel data storage, neuromorphic computation and multi-level memories.

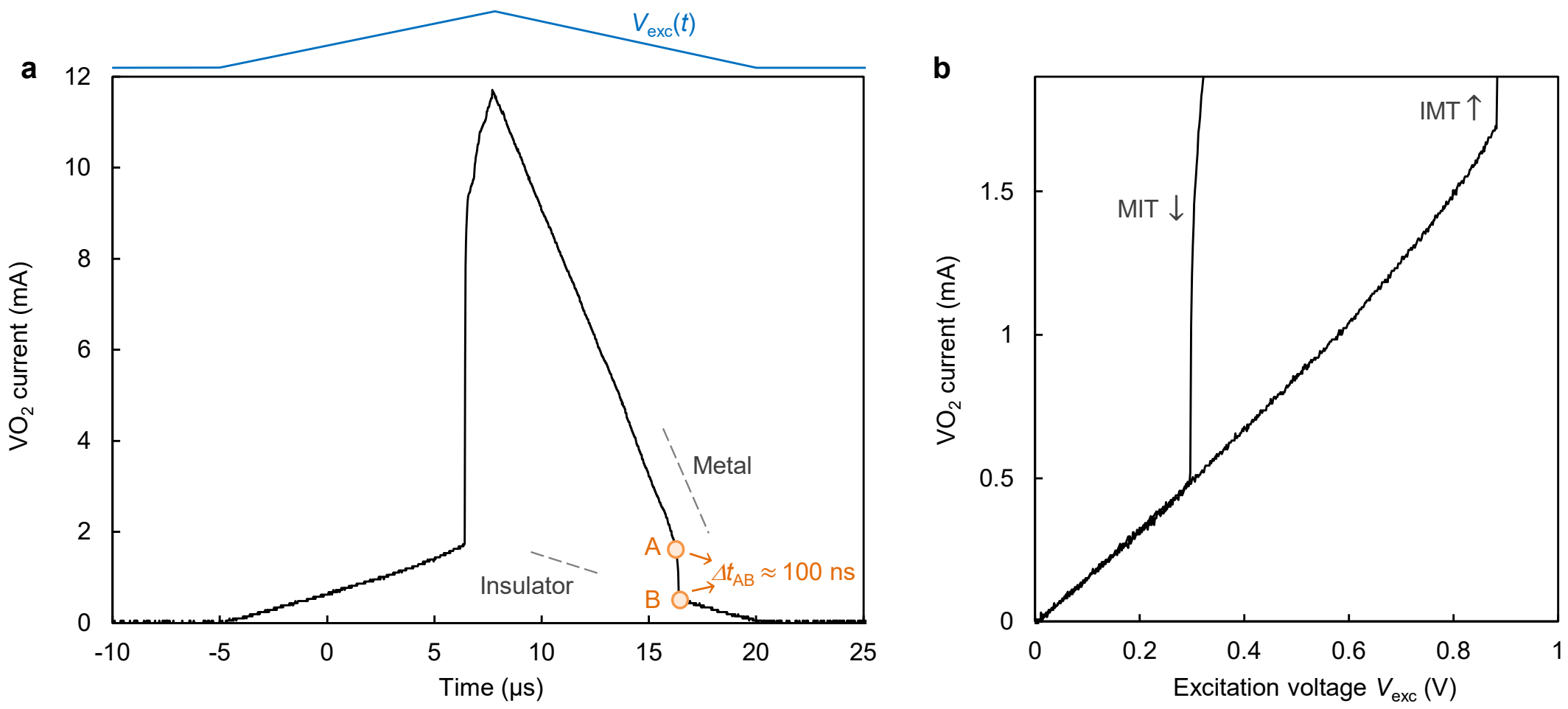

**Fig. S1. Transient current-voltage characterization showing a fast cooling rate.** (**a**) Measured current of a $VO_2$ switch under triangular excitation. The MIT transition happens between points A and B with a short time separation of $\Delta t_{AB} \approx 100$ ns. (**b**) Extracted resistance of the device in the IMT and MIT cycles showing that post excitation resistance is close (within 1%) of the pre-IMT resistance which is another indication of fast cooling.

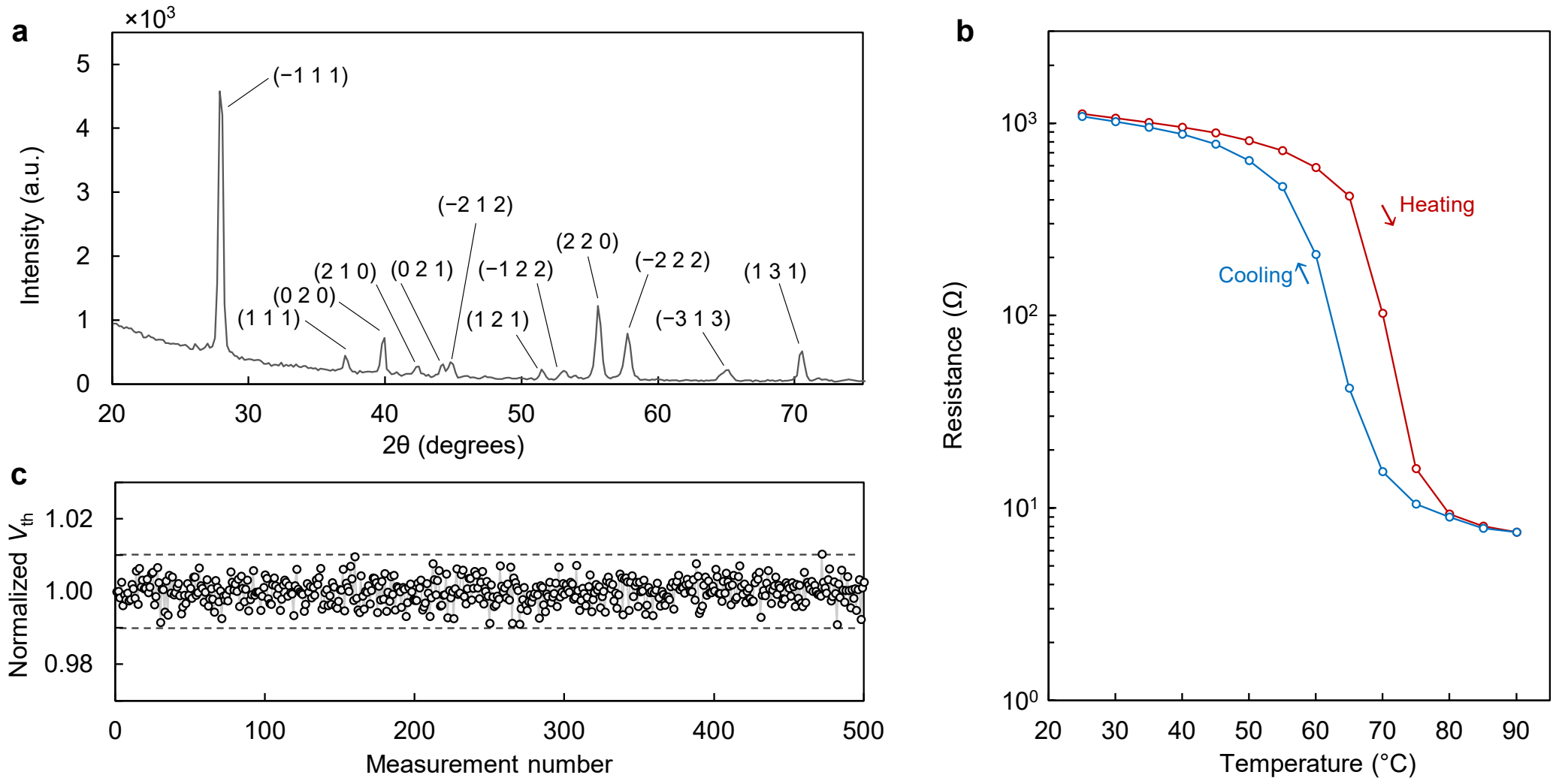

**Fig. S2. Evaluation of Vanadium Dioxide film and fabricated devices.** (**a**) The thin film structure was determined by X-ray diffraction (XRD) analysis using an Empyrean X-ray diffractometer with monochromatic Cu Kα radiation (λ=0.154056 nm). The θ − 2θ diffraction pattern, recorded in the 20° – 75° (2*θ*) range. The result indicates that the sample is crystallized in the single monoclinic phase as all diffraction peaks are indexed to monoclinic $VO_2$ (M1) crystal structure according to PDF 04-003-4401 (Space group $P2_1/c$, a = 5.75 Å, b = 4.52 Å, c = 5.38 Å, β = 122.6°). (**b**) Resistance versus temperature of a two-port $VO_2$ switch showing a sharp reduction at the critical temperature of $VO_2$. (**c**) Threshold consistency indicating no degradation. A 3-μm-long $VO_2$ switch was excited by a voltage ramp and the threshold voltage at each IMT triggering was monitored. The results indicate less than ±1% variation with no drift, indicating no degradation in the film.

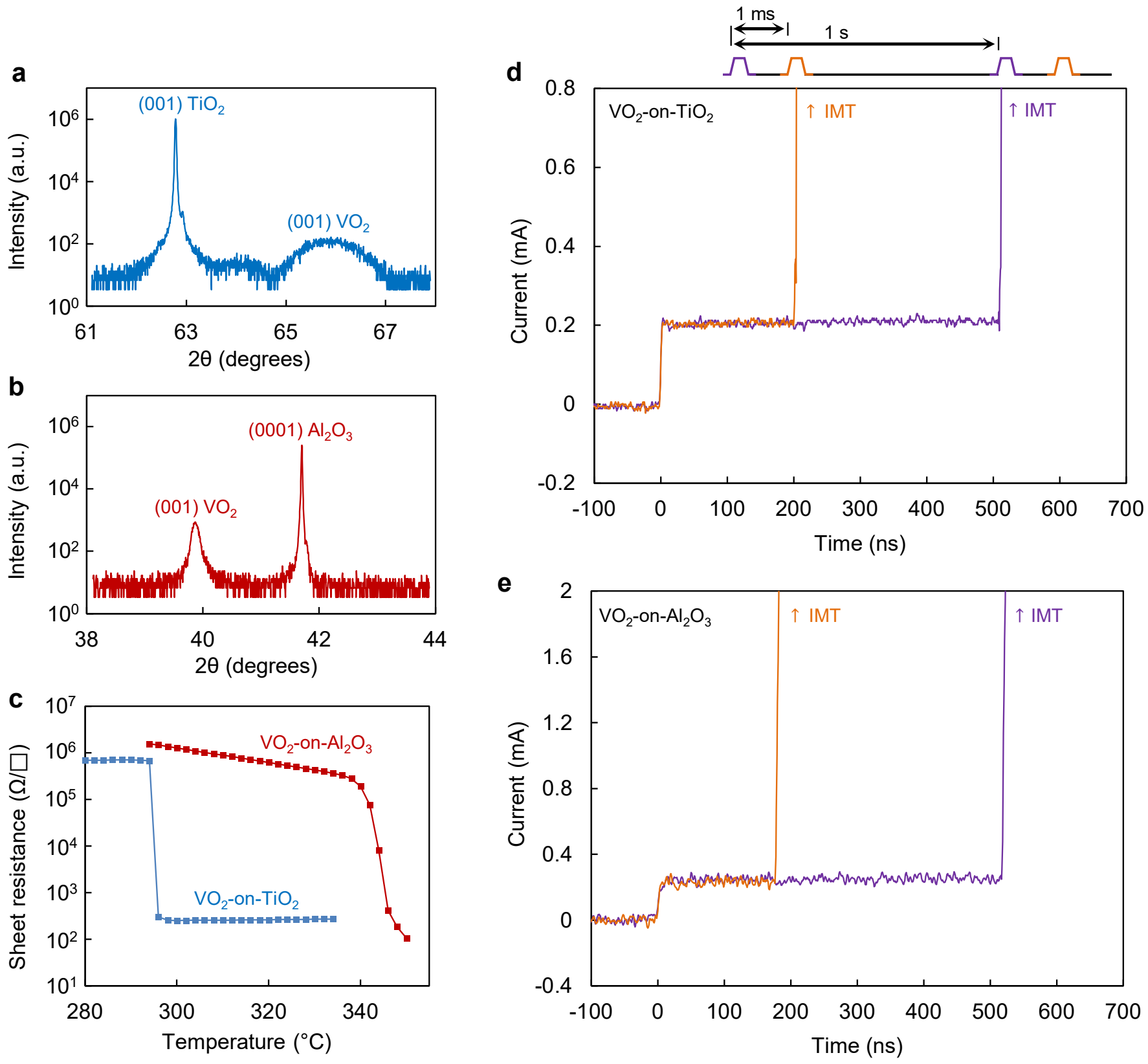

**Fig. S3. Reproducibility of the results on a single-crystal $VO_2$ samples on $TiO_2$ and $Al_2O_3$ substrates.** Symmetrical 2θ-θ XRD scan on the (**a**) 10 nm-thick $VO_2$ film grown on (001) $TiO_2$ substrate ($VO_2$-on-$TiO_2$), and (**b**) 100 nm thick $VO_2$ film grown on (0001) $Al_2O_3$ ($VO_2$-on-$Al_2O_3$). (**c**) Sheet resistance measurements on $VO_2$-on-$TiO_2$ and $VO_2$-on-$Al_2O_3$ samples in the heating cycle. Pulsed-measurements with two relaxation times 1-ms and 1-s on (**d**) $VO_2$-on-$TiO_2$ and (**e**) $VO_2$-on-$Al_2O_3$ samples, showing the strong dependence of $t_{inc}$ on the relaxation time.

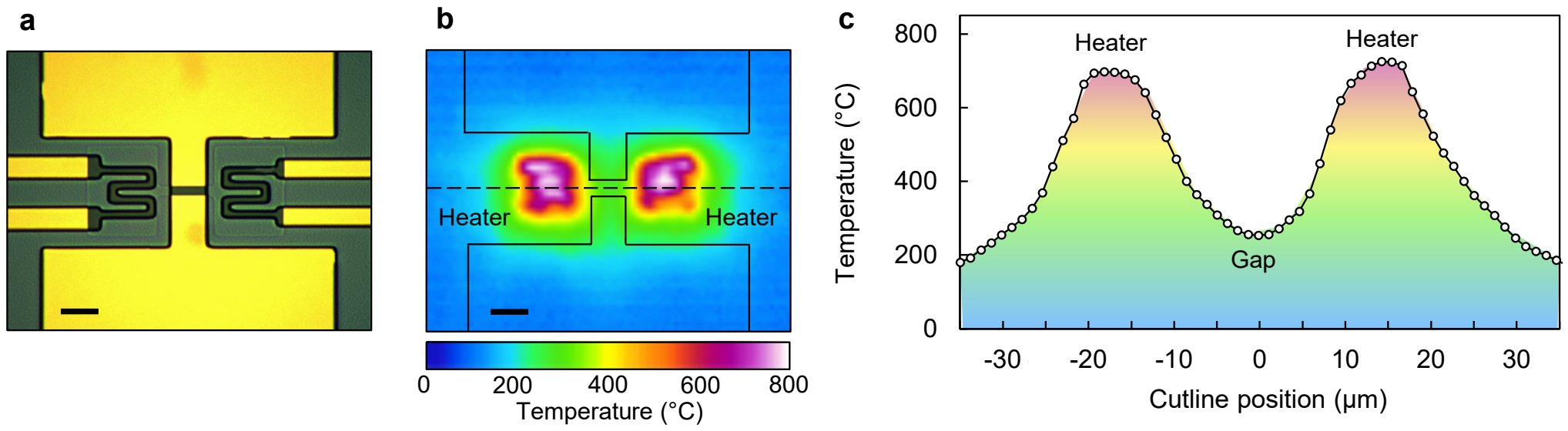

**Fig. S4. Thermal activation of IMT using isolated heaters.** (**a**) Optical micrograph of a $VO_2$ switch integrated with two electrically isolated heaters. The scale bar corresponds to 10 µm. (**b**) Captured thermal micrograph of the device with the heaters are triggered at their threshold voltage (18 V). The scale bar corresponds to 10 µm. (**c**) Measured temperature over the cutline shown in part B indicating that the temperature at the middle switch (gap) considerably surpasses the IMT temperature. The measurement is done at the steady state, however, considering a sub-microsecond thermal time constant (Fig. S1a), for a 20-µs long excitation on the heaters, a thermal IMT in the middle switch is expected.

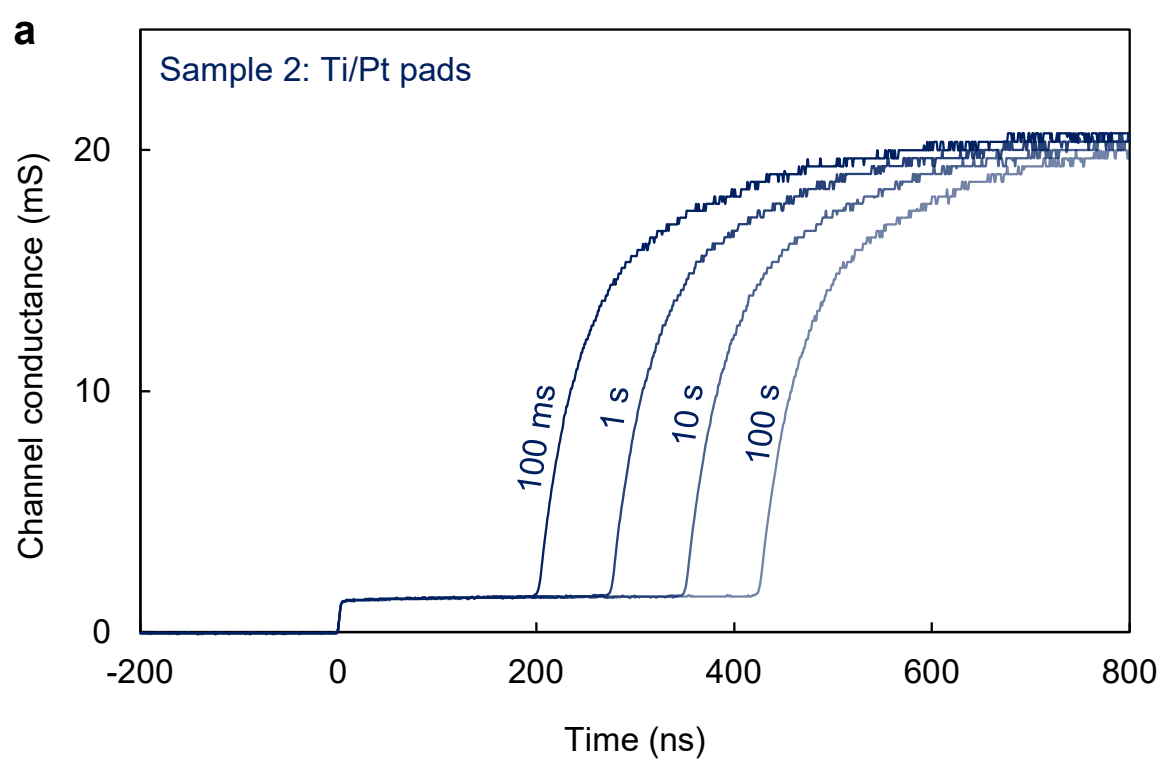

**Fig. S5. Reproducibility of the results in devices with other metallic pads.** Two-port switches fabricated on the high resistivity silicon substrate, based on (**A**) $VO_2$ / Ti (10 nm) / Pt (200 nm) and (**B**) $VO_2$ / Pt (200 nm) structures exhibit identical memory behavior. The results indicate that the observation is independent from metal-$VO_2$ interface.

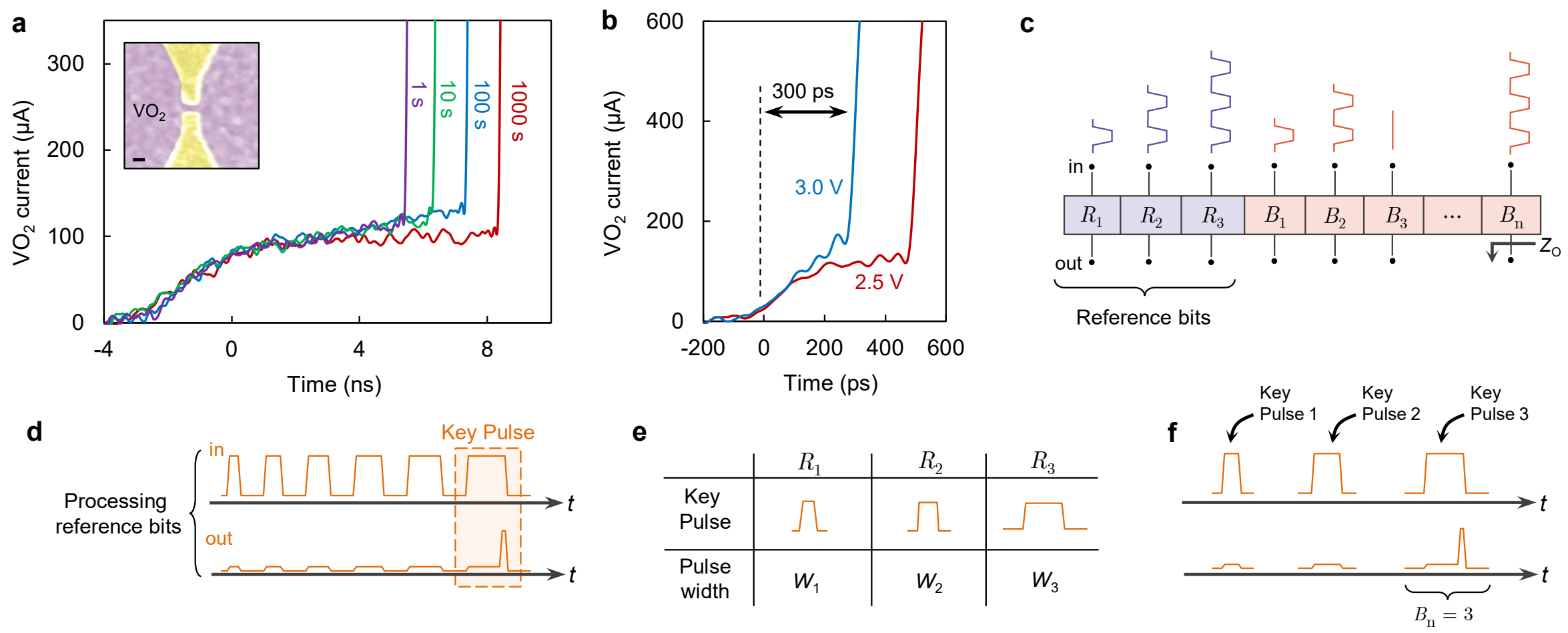

**Fig. S6. Concept of ultrafast ultra-scaled multilevel Mott-based memory.** (**a**) Observation of the memory effect in a 50-nm-gap device. Measured $VO_2$ current is shown for different pulse separations of 1 s, 10 s, 100 s, and 1000 s. The device was excited by a 50-Ω port of pulse generator were the amplitude was set to 2 V. The device exhibited switching in $\lesssim$ 10 ns. The inset shows the false colored scanning electron microscopy (SEM) image of the device. The scale bar is 100 nm. (**b**) Demonstration of an ultrafast sub-nanosecond writing with a low energy level ~0.1 pJ. It should be noted that the mesa region of the device has not been etched (3a, inset), and so a considerable part of the pre-transition current is the fringing current which does not directly contribute on IMT triggering in the gap. As a result, the writing energy can be future scaled down by an optimized device layout. (**c**) An example scheme of four-level Mott-based memory including three reference bits $R_1$, $R_2$, and $R_3$ (written by different number of pulses, $N_1$, $N_2$, and $N_3$, respectively) and $n$ ordinary bit $B_k$. (**d**) Reading process starts by processing reference bits to extract the pulse width needed to trigger IMT in each reference bit. (**e**) Three key pulses with pulse-widths $W_1$, $W_2$, and $W_3$ are obtained ($W_1 < W_2 < W_3$). (**f**) Three key pulses are applied to each bit $B_k$ to accomplish the reading process.